\documentclass[12pt]{article}
\let\section=\subsection     \let\subsection=\subsubsection                
\usepackage{graphicx}

\newcommand{\expl}{\langle \!\langle}
\newcommand{\expr}{\rangle \!\rangle}
\def\Journal#1#2#3#4{{#1} {\bf #2}, #3 (#4)}

\def\NPA{{\em Nucl.~Phys.} A}
\def\PLB{{\em Phys.~Lett.}  B}
\def\PRL{\em Phys.~Rev.~Lett.~}

\def\PRC{{\em Phys.~Rev.} C}
\def\ZPC{{\em Z.~Phys.} A}
\def\ZPC{{\em Z.~Phys.} C}

\def\PRep{\em Phys.~Rep.~}

\def\JP{{\em J.~Phys.} G}

\begin{document}
\begin{center}
   {\large \bf CHEMICAL EQUILIBRATION OF ANTIHYPERONS}\\[5mm]
   C.~GREINER \\[5mm]
   {\small \it  Institut f\"ur Theoretische Physik, Universit\"at Giessen, \\
   Heinrich-Buff-Ring 16, D-35392 Giessen, Germany \\[8mm] }
\end{center}
\begin{abstract}\noindent
Rapid chemical equilibration of antihyperons
by means of the interplay between strong annihilation on baryons
and the corresponding backreactions of multi-mesonic (fusion-type)
processes in the later, hadronic
stage of an ultrarelativistic heavy ion collision will be discussed.
Explicit rate calculations for a dynamical setup are presented.
At maximum SPS energies yields
of each antihyperon specie are obtained which are consistent
with chemical saturated populations of $T \approx  150-160 $ MeV.
The proposed picture supports dynamically the popular
chemical freeze-out parameters extracted within thermal models.
\end{abstract}

\section{Physics of antihyperon production}

Strangeness enhancement
has been predicted  a long time ago as a
diagnostic probe to prove for the short-time existence of a
QGP in the course of a relativistic
heavy ion collision.
With respect to the high production thresholds in the
various binary hadronic reaction channels,
especially the antihyperons were then advocated as the
appropriate candidates \cite{KMR86}.
Assuming the
existence of a temporarily present phase of QGP and
following statistical coalescence estimates,
the abundant (anti-)strange quarks
can easily be redistributed to combine
to strange (anti-)baryons \cite{KMR86}, which do then,
in return, come close to their chemical equilibrium values
at the onset of the hadronic phase.

Indeed, such a behaviour of nearly chemically saturated populations of the
antihyperons has been experimentally demonstrated
with the Pb+Pb experiments NA49 and WA97 at CERN-SPS.
For this statement, of course, a quantitative,
theoretical analysis within a thermal model
has to be invoked by fitting the thermodynamical parameters
to the set of individual (strange) hadronic abundancies \cite{BMS96}.
Some phenomenological attempts
to explain a more abundant production of antihyperons within
a hadronic transport description \cite{So95} do exist
like the color rope or the high-dense cluster formation.
The underlying mechanisms, however, have to be considered as exotic.
In particular, a dramatic role of antibaryon annihilation
is observed, which, in return, has to be more than counterbalanced
in a rather ad hoc way by these more exotic mechanisms.

Be it as it is, a correct incorporation of the strong baryonic annihilation
channels had actually not been achieved within the various microscopic
implementations.
Recently we have conjectured that
a sufficiently fast redistributions of strange and light quarks
into (strange) baryon-antibaryon pairs should be achieved
by multi-mesonic fusion-type reactions of the type
\begin{equation}
\label{mesfuse}
n_1\pi + n_2 K \, \leftrightarrow \, \bar{Y}+p
\end{equation}
for a moderately dense hadronic system \cite{GL00}.
The beauty of this argument lies in the fact that
(at least) these special kind of multi-hadronic
reactions have to be present because of the fundamental principle of detailed
balance. As the annihilation of antihyperons on baryons is of dramatic relevance,
the multi-mesonic (fusion-like) `back-reactions'
involving $n_1$ pions and $n_2$ kaons, where $n_2$ counts
the number of anti-strange quarks within the antihyperon $\bar{Y}$),
must, in principle, be taken care of in a dynamical simulation.
This reasoning has first been raised by Rapp and Shuryak
concerning the production of anti-protons \cite{RS00}, which then
triggered our work.

It is plausible to assume that the
annihilation crosssections are approximately the same as for $N\bar{p}$
at the same relative momenta, i.e.
$\sigma _{B \bar{Y}\rightarrow n  \pi + n_Y K} \equiv \sigma _0 \approx 50 $ mb.
The equilibration timescale
$(\Gamma _{\bar{Y}})^{(-1)} \sim
1/ (\sigma _{B \bar{Y}} v _{\bar{B}N} \rho_B )$
is to a good approximation proportional to the
inverse of the density of baryons and their resonances.
Adopting an initial density of 1--2 times
normal nuclear matter density $\rho_0 $ for the initial and thermalized
hadronic fireball, the antihyperons will equilibrate
on a timescale of 1--3 fm/c! This timescale competes
with the expansion timescale of the late hadronic fireball.
To be more quantitative rate calculations for a dynamical setup will be presented
in the next section.

\section{Rate equation results and their implications}

Following the concepts of kinetic theory, the
microscopic starting point would be
a Boltzmann-type equation of the form
{\footnotesize
\begin{eqnarray}
\partial_t f_{\bar{Y}} + \frac{{\bf p}}{E_{\bar{Y}}} {\bf \nabla } f_{\bar{Y}}
& = &
\sum_{ \{ n_1 \} ; B} \frac{1}{2E_{\bar{Y}}}
\int \frac{d^3p_B}{(2\pi )^3 2E_B}
\prod _{ \{ n_1, n_2 \} } \int \frac{d^3p_i}{(2\pi )^3 2E_i}
\, (2\pi )^4
\delta^4 ( p_{\bar{Y}} + p_B - \sum_{\{ n_1, n_2 \} }p_i )
\nonumber \\
&&
| \expl n_1,n_2 | T | \bar{Y} B \expr |^2
\left\{ (-) f_{\bar{Y}}f_B \prod_{\{n_1, n_2 \} } (1+f_i) \, + \,
\prod _{ \{ n_1, n_2 \} } f_i
(1- f_{\bar{Y}}) (1-f_B) \right\} \, \, ,
\nonumber
\end{eqnarray}
}
where
{\footnotesize
\begin{eqnarray}
\label{cross}
\sigma _{\bar{Y}B}^{\{n_1 \} } & \equiv &
\frac{1}{`Flux'}
\prod _{ \{ n_1, n_2 \} } \int \frac{d^3p_i}{(2\pi )^3 2E_i}
\, (2\pi )^4
\delta^4 \left( p_{\bar{Y}} + p_B - \sum_{ \{ n_1, n_2 \} }p_i \right)
| \expl n_1,n_2 | T | \bar{Y} B \expr |^2
\nonumber
\end{eqnarray}
}
corresponds to the annihilation cross section.
Assuming $v_{rel}\, \sigma_{\bar{Y}B} (\sqrt{s} ) $ to be roughly constant,
which is actually
a good approximation for the $p\bar{p}$-annihilation,
or, invoking a standard coarse grained description of
thermally averaged cross sections
and distributions,
and furthermore
taking the distributions in the Boltzmann approximantion,
the following master equation
for the respectively considered antihyperon density
is obtained
\begin{equation}
\label{mastera}
\frac{d}{dt} \rho _{\bar{Y}} \, =\,   -
\expl \sigma _{\bar{Y}B} v _{\bar{Y}B} \expr
\left\{
\rho _{\bar{Y}} \rho_B \,  \vphantom{\sum_{n}}
-  \, \sum_{\{ n_1 \} }
{\cal R}_{(n_1,n_2)}(T,\mu_B,\mu_s) (\rho _\pi)^{{n_1}} (\rho _K )^{n_2}
\right\},
\end{equation}
where the
`back-reactions' of several effectively coalescing
pions and kaons are incorporated in the `mass-law' factor
$$
{\cal R}_{(n,n_Y)}(T,\mu_B,\mu_s) \, = \,
\frac{ \rho _{\bar{Y}}^{eq.} \rho ^{eq.}_B }
{(\rho ^{eq.}_\pi)^{{n_1}} (\rho ^{eq.}_K )^{n_2}} \, p_{n_1} \, \, \, .
$$
Here $p_{n_1}$ states the relative probability
of the reaction (\ref{mesfuse}) to decay into a specific number $n_1$ of pions
and $\rho_B$ denotes the total number density of baryonic particles.
${\cal R}$ depends only on the temperature and the baryon and strange quark chemical
potentials.
$\Gamma  _{\bar{Y}} \equiv
\expl \sigma _{\bar{Y}N} v _{\bar{Y}N} \expr \rho_B $
gives the effective annihilation rate of the respective antihyperon
specie on a baryon.

Nonequilibrium inelastic hadronic reactions
can explain to a good extent the overall strangeness production
seen experimentally:
The major amount of the produced kaons at SPS-energies
can be understood in terms of still early and energetic
non-equilibrium interactions \cite{Ge98,CG01}.
Refering to the master equation (\ref{mastera}),
one can then take the pions, baryons and kaons to stay approximately
in thermal equilibrium throughout the later hadronic evolution of the collision,
the later being modelled to be an isentropic expansion
with fixed total entropy content being
specified via the entropy per baryon ratio $S/A$.
(\ref{mastera}) becomes
\begin{equation}
\label{masterd}
\frac{d}{dt} \rho _{\bar{Y}} \,  = \,    - \,
\Gamma  _{\bar{Y}}
\left\{
\rho _{\bar{Y}} \, -  \,
\rho ^{eq }_{\bar{Y}}
\right\} \, \, \, .
\end{equation}

The `effective' (global or at midrapidity) volume $V(t)$
is parametrized as function of time by longitudinal
Bjorken expansion and including an accelerating
radial flow, e.g.
\begin{equation}
\label{volume}
V_{eff}(t\geq t_0) \, = \, \pi \, (ct) \,
\left(R_0 + v_0(t-t_0) + 0.5 a_0 (t-t_0)^2 \right)^2
\end{equation}
with $R_0 = 6.5\, fm$, $v_0= 0.15\, c$ and $a_0 = 0.05 \, c^2/fm$.
At starting time $t_0 $ an initial temperature $T_0$ is chosen.
($T_0$ is set to $190$ MeV for the SPS
and $150 MeV$ for the AGS situation, while
the initial energy densities are
then about 1 GeV/fm$^3$.)
From (\ref{volume}) together with the constraint of conserved entropy
the temperature and the chemical potentials do
follow as function of time.
As a {\em minimal } assumption the initial abundancy of antihyperons
is set to zero. Equation (\ref{masterd}) is solved for each specie.

\begin{center}
\vspace*{-2mm}
   \includegraphics[height=7cm]{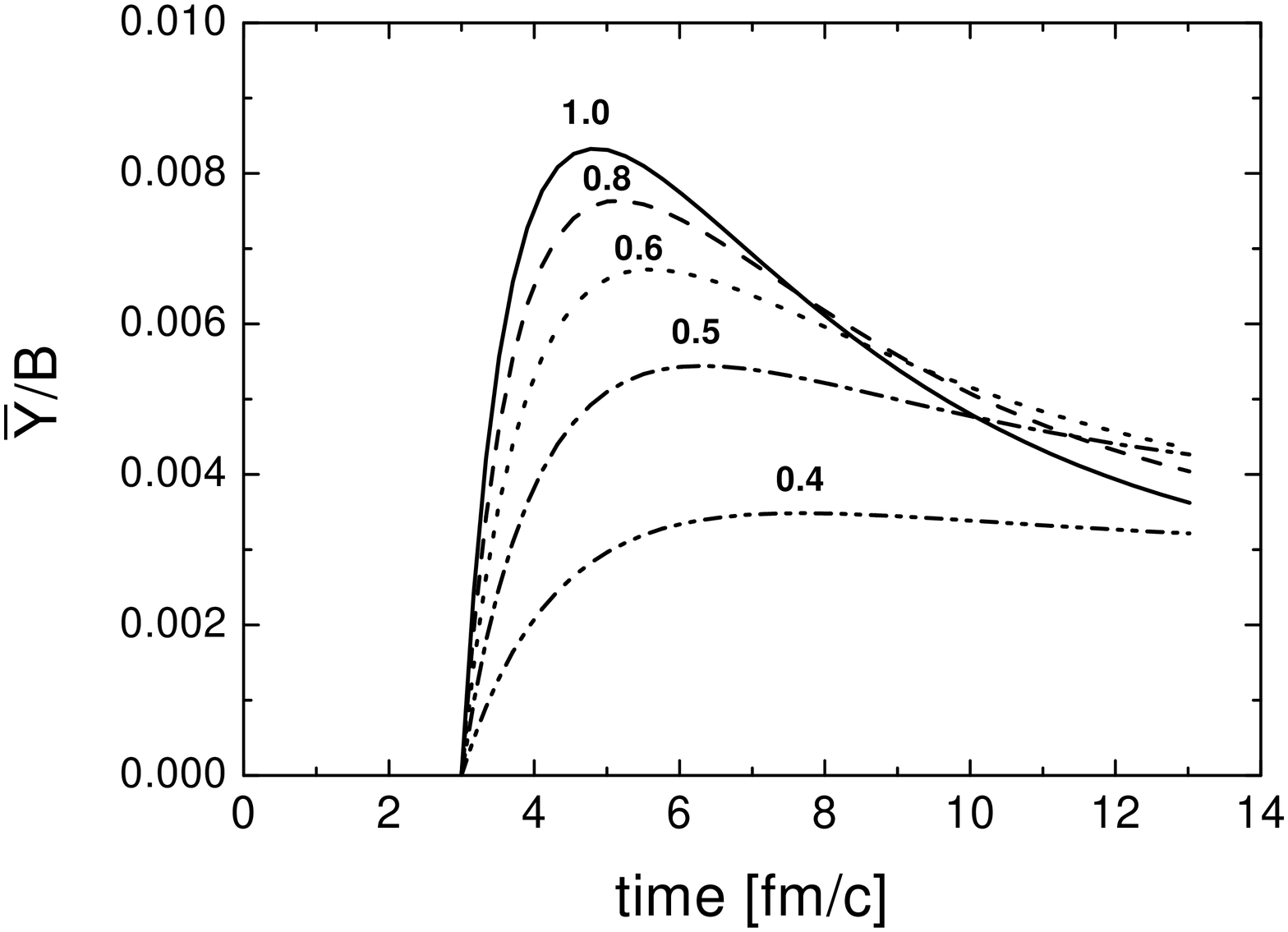}\\
   \parbox{14cm}
        {\footnotesize
        Fig.~1: The anti-$\Lambda $ to baryon number ratio
$N_{\bar{ \Lambda }}/N_B \, (t)$
as a function of time for various
implemented annihiation cross section $\sigma_{eff} \equiv \lambda \,
\sigma_0 $.
The entropy per baryon is taken as $S/A=30$, $t_0=3 $ fm/c and
$T_0=190 $ MeV.
}
\end{center}

In Fig.~1 the number of $\bar{\Lambda }$s
(normalized to the conserved net baryon number)
as a function of time is depicted.
The entropy per baryon is chosen as $S/A=30$ being characteristic
to global SPS results. In addition the cross section employed
is varied by a constant factor, i.e.
$\sigma_{eff} \equiv \lambda \, \sigma_0$.
The characteristics is that first the antihyperons
are dramatically being poulated, and then in the very late
expansion some more are still being annihilated.
The results are rather robust against a variation by a factor of 2
in the cross section.

In Fig.~2 the number of antihyperons of each specie
are now shown as a function of the decreasing temperature $T(t)$
of the hadronic system.
For a direct comparison the instantaneous equilibrium abundancy
$N_{\bar{Y}}^{eq.}(T(t),\mu_B(t),\mu_s(t))/N_B$ is also given.
As noted above, after a fast initial population,
the individual yields of the antihyperons do overshoot
their respective equilibrium number
and then do finally saturate at some slightly smaller value.
Moreover, one notices that
the yields effectively do saturate at a number
which can be compared to an equivalent equilibrium number
at a temperature parameter around $T_{eff} \approx 150-160$ MeV,
being strikingly close
to the ones obtained within the various thermal analyses \cite{BMS96}.

\begin{center}
\vspace*{-2mm}
   \includegraphics[height=8cm]{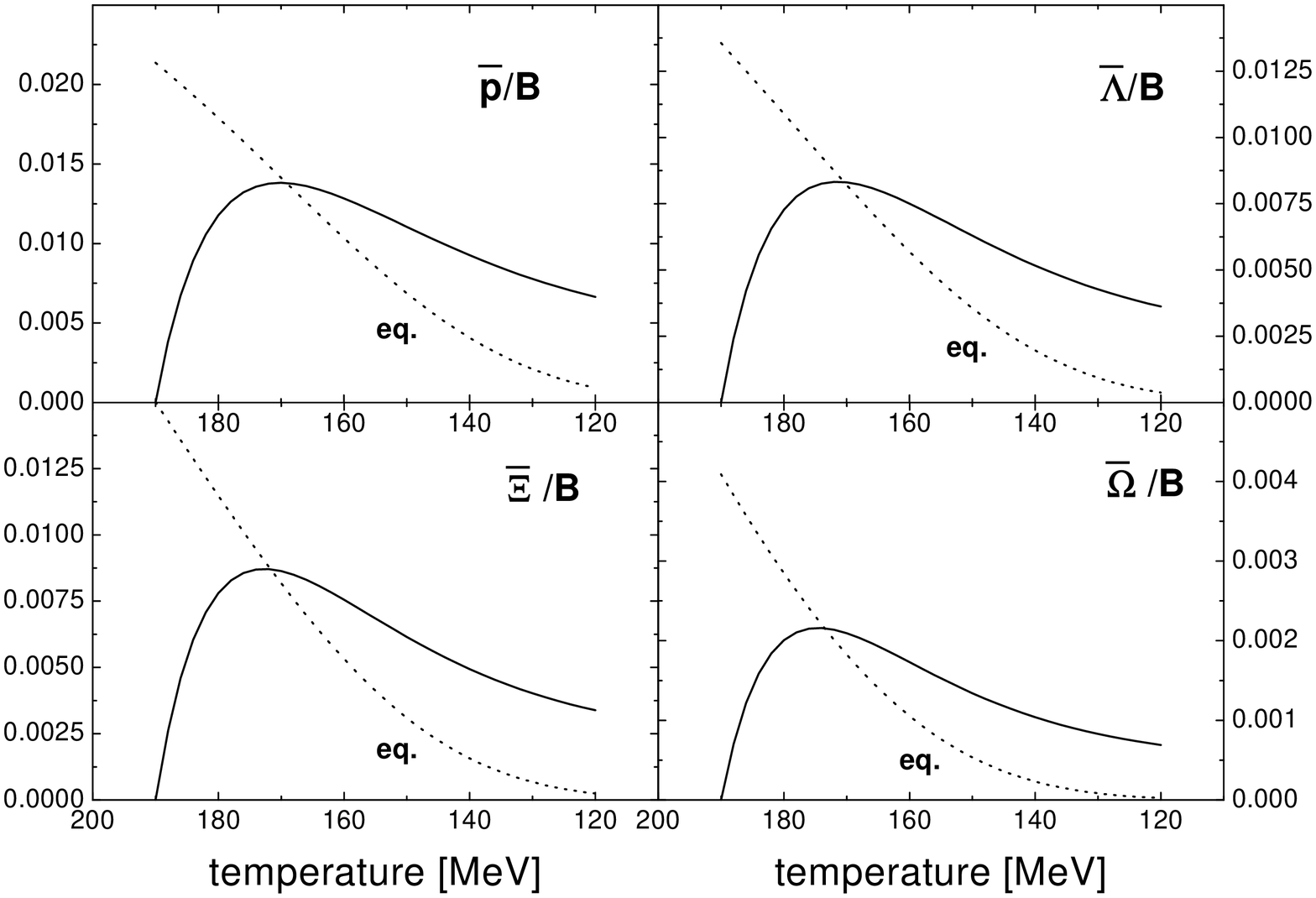}\\
   \parbox{14cm}
        {\footnotesize
  Fig.~2: The antihyperon to baryon number ratio
$N_{\bar{Y}}/N_B \,  (T)$ and $N_{\bar{Y}}^{eq.}/N_B (T) \,  $
(dotted line) as a function of the decreasing temperature.
Parameters are the same as in Fig. 1.
}
\end{center}

In Fig.~3 the number of anti-$\Lambda $s
as a function of time is given for various entropy per baryon ratios.
One notices that the final value in the yield significantly
depends on the entropy content, or, in other
words, on the baryochemical potential. We note that the results
at midrapidity from WA97 can best be reproduced by employing
an entropy to baryon ratio $S/A=40$, where one qualitatively
expects a larger entropy content.

There is also a clear hint at AGS energies
of enhanced anti-$\Lambda $ production \cite{E802}.
For most central collisions more anti-$\Lambda $s are found
compared to anti-protons, which is quite puzzling as within
a thermal model analysis this ratio is found not to be larger
than 1. This enhanced ratio of anti-$\Lambda $s compared to
anti-protons at AGS energies one can
understand in a way that one assumes that their annihilation cross section
on baryons is just slightly smaller than for the antiprotons.
In Fig.~4 a similar study like that of Fig.~1 is shown for a characteristic
situation at AGS. For smaller, yet not too small effective cross sections
the final yield can here be enlarged by a factor of 2 compared to the
case with a `full' crossection, as the final reabsorption is not as effective.
But, of course, this idea is speculation at present.
Also, we remark that
the $\bar{\Lambda }$s effectively do saturate
at an equivalent equilibrium number
at a temperature parameter around $T_{eff} \approx 120-130$ MeV.
Unfortunately, there are no data for $\bar{\Xi }$ at AGS.
A detailed measurement of all antihyperons
represents an excellent opportunity for future heavy ion facilities
at an energy upgraded GSI.

\begin{center}
\vspace*{-2mm}
   \includegraphics[height=60mm,width=68mm]{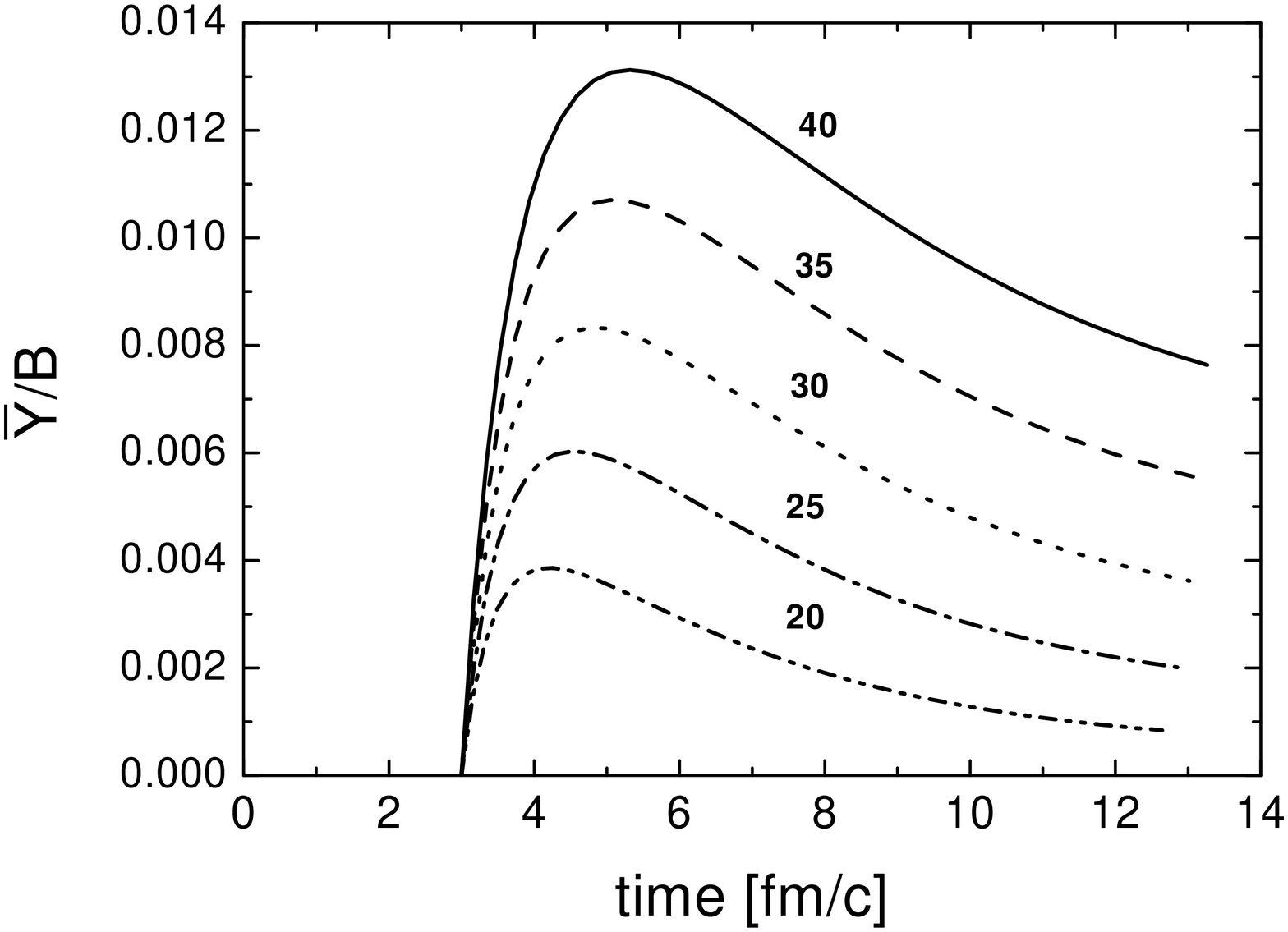}
   \hspace{\fill}
   \includegraphics[height=60mm,width=68mm]{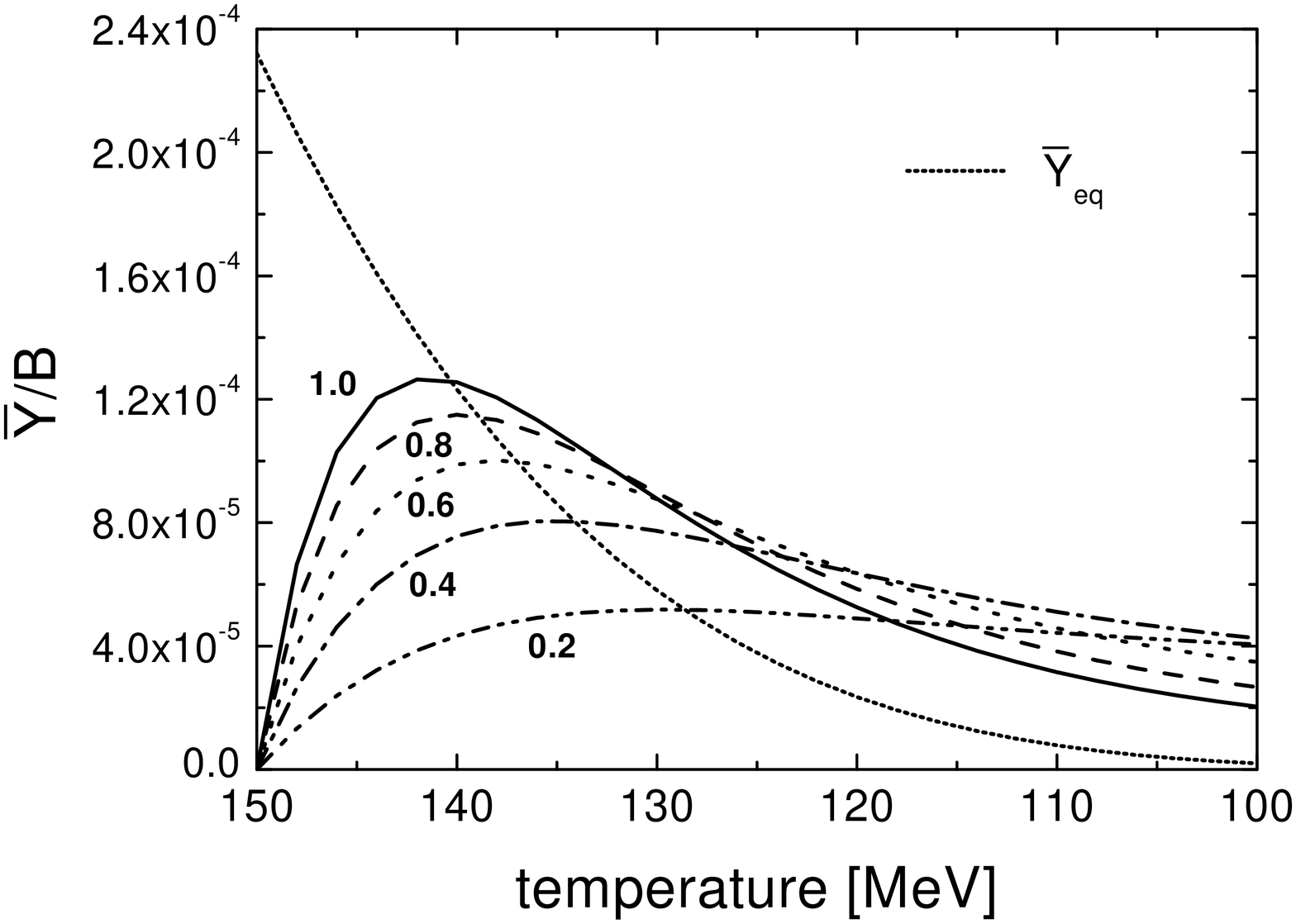} \\
   \parbox{68mm}
        {\footnotesize
      Fig.~3: $N_{\bar{\Lambda }}/N_B \,(t) $ as a function of time
     for various entropy content described
  via the entropy per baryon ratio ($S/A=20-40$).
  Other parameters are as in Fig.~1.}
   \hspace{\fill}
   \parbox{68mm}
        {\footnotesize
      Fig.~4: $N_{\bar{\Lambda }}/N_B \,(T) $
      and $N_{\bar{\Lambda }}^{eq.}/N_B (T) \,  $
      as a function of
      decreasing temperature for a characteristic AGS situation
     with an entropy content of $S/A=12$ for various
    implemented annihiation cross section
    $\sigma_{eff} \equiv \lambda \, \sigma_0 $.
   $t_0=5 $ fm/c and
   $T_0=150 $ MeV. }
\end{center}

\end{document}